\documentclass[oneculumn]{aastex631}
\usepackage{xspace}
\usepackage{graphicx}

\newcommand\h     {$^{\rm h}$}
\newcommand\m     {$^{\rm m}$}
\newcommand\s     {$^{\rm s}$}
\newcommand{\solmass}  {M$_{\odot}$}
\newcommand{\kms} {\ifmmode{\rm \,km\,s^{-1}}\else\,km\,s$^{-1}$\xspace\fi}

\newcommand{\ha}{\hbox{H$\alpha$}}
\newcommand{\hb}{\hbox{H$\beta$}}
\newcommand{\hg}{\hbox{H$\gamma$}}
\newcommand{\sii}{\hbox{[S\,{\sc ii}]}}
\newcommand{\nii}{\hbox{[N\,{\sc ii}]}}
\newcommand{\oiii}{\hbox{[O\,{\sc iii}]}}
\newcommand{\hii}{\hbox{H\,{\sc ii}}}
\definecolor{mycolor}{rgb}{0.858, 0.188, 0.478}

\shortauthors{Chung et al.}

\accepted{October 13, 2021}
\submitjournal{ApJ}

\begin{document}

\title{Star-forming Dwarf Galaxies in Filamentary Structures around the Virgo Cluster: Probing Chemical Pre-processing in Filament Environments}

\author[0000-0003-0469-345X]{Jiwon Chung}
\affil{Korea Astronomy and Space Science Institute 776, Daedeokdae-ro, Yuseong-gu, Daejeon 34055, Republic of Korea; jiwon@kasi.re.kr}

\author[0000-0002-3738-885X]{Suk Kim}
\affil{Department of Astronomy and Space Science, Chungnam National University, Daejeon 34134, Republic of Korea}
\author[0000-0002-0041-6490]{Soo-Chang Rey}
\affil{Department of Astronomy and Space Science, Chungnam National University, Daejeon 34134, Republic of Korea; screy@cnu.ac.kr}
\author[0000-0002-6261-1531]{Youngdae Lee}
\affil{Department of Astronomy and Space Science, Chungnam National University, Daejeon 34134, Republic of Korea}

\begin{abstract}
It has been proposed that the filament environment is closely connected to the pre-processing of galaxies, where their properties may have been changed by environmental effects in the filament before they fell into the galaxy cluster. We present the chemical properties of star-forming dwarf galaxies (SFDGs) in five filamentary structures (Virgo III, Leo Minor, Leo II A, Leo II B, and Canes Venatici) around the Virgo cluster using the Sloan Digital Sky Survey optical spectroscopic data and Galaxy Evolution Explorer ultraviolet photometric data. We investigate the relationship between stellar mass, gas-phase metallicity, and specific star formation rate (sSFR) of SFDGs in the Virgo filaments in comparison to those in the Virgo cluster and field. We find that, at a given stellar mass, SFDGs in the Virgo filaments show lower metallicity and higher sSFR than those in the Virgo cluster on average. We observe that SFDGs in the Virgo III filament show enhanced metallicities and suppressed star formation activities comparable to those in the Virgo cluster, whereas SFDGs in the other four filaments exhibit similar properties to the field counterparts. Moreover, about half of the galaxies in the Virgo III filament are found to be morphologically transitional dwarf galaxies that are supposed to be on the way to transforming into quiescent dwarf early-type galaxies. Based on the analysis of the galaxy perturbation parameter, we propose that the local environment represented by the galaxy interactions might be responsible for the contrasting features in "chemical pre-processing" found in the Virgo filaments.

\end{abstract}



\section{Introduction} \label{sec:intro}

The large-scale structure of the Universe is characterized by a web-like network composed of filaments that intersect at nodes wherein clusters of galaxies are found \citep{Bond1996}. Filaments are the most dominant structures in the distribution of galaxies, extending over scales up to tens of Mpcs. Moreover, a significant fraction of the mass in the Universe is concentrated in filaments \citep{Aragon-Calvo2010}. The hierarchical models of structure formation predict that galaxy clusters grow by the continuous accretion of galaxies through filaments \citep{Ebeling2004}. 

The role of the filament on the properties and evolution of galaxies has been conducted by numerous studies regarding the subject of "pre-processing" \citep{Fujita2004}. In particular, there is a growing body of observational results on star formation quenching of galaxies in the filament environments \citep{Martinez2016,Kuutma2017,Castignani2021}). The changes of galaxy properties as a function of the distance from the filament spine or node have also been extensively investigated \citep{Alpaslan2016,Martinez2016,Chen2017,Kuutma2017,Malavasi2017,Kraljic2018,Laigle2018,Mahajan2018,Luber2019,Sarron2019,Bonjean2020,Lee2021}. On the other hand, several studies have also reported that the filament environment connected with the cluster is responsible for enhanced star formation activity and an increased fraction of star-forming galaxies in the filaments \citep{Porter2007,Fadda2008,Biviano2011,Coppin2012,Darvish2014}. These observational results  provide evidence that the properties of galaxies may have been changed in the filaments via several environmental effects before they fell into galaxy clusters.   

From the perspective of the provision of invaluable information on galaxy evolution, gas-phase metallicity could be a unique observable parameter for probing the chemical evolution of galaxies. As galaxies evolve, metals produced by nucleosynthesis are released into the surrounding interstellar medium during the later evolutionary stages of stars. As the cycle of star formation recurs, the metals in the interstellar medium of the galaxy become steadily enriched. Thus, the metallicity traces a fossil record of the star formation history.    
Several studies have also attempted to identify the relationship between the chemical properties of galaxies and their surrounding environments (e.g., \citealt{Skillman1996,Pilyugin2002,Hughes2013}). \citet{Skillman1996} and \citet{Pilyugin2002} found that spiral galaxies near the core of the Virgo cluster have higher oxygen abundances than those in the field environment. However, \citet{Hughes2013} suggested that internal evolutionary processes, rather than environmental effects, play a key role in the chemical evolution of cluster galaxies. While the role of the environment in shaping the chemical evolution of galaxies remains a controversial issue, a sophisticated approach that can provide accurate understandings of the impact of the environment is required.

To date, observational studies of filament galaxies using gas-phase metallicity have focused on massive ($>$ 10$^{9}$ \(M_\odot\)), bright galaxies mainly because of their affluence of observations (e.g., \citealt{Darvish2015}). However, low-mass galaxies are more vulnerable to even weak external influences, due to their shallow gravitational potential wells, compared to that of massive galaxies. Thus, they are ideal objects for probing various processes that alter the properties and evolution of galaxies in filament environments.

The Virgo cluster is the nearest rich cluster from the Milky Way at a distance of $\sim$17 Mpc \citep{Mei2007,Boselli2014}. Moreover, since the Virgo cluster is considered to be a dynamically young cluster \citep{Aguerri2005}, a connection of the galaxy distribution of the cluster to the nearby filament structures was expected \citep{Tully1982}. Recently, \citet{Kim2016} defined the filament structures that are physically connected to the Virgo cluster. These Virgo filaments are mostly composed of faint dwarf galaxies (M$_B$ $>$ $-$19 mag; $\sim$88$\%$ of the total sample) and exhibit a range of characteristics (e.g., galaxy number density, length, and thickness of the filaments; \citealt{Lee2021}). Most recently, two studies have quantified the properties of galaxies in the Virgo filaments, in terms of their colors, star formation rate, and gas content with respect to the vertical distance from the filament spine and local galaxy density \citep{Lee2021,Castignani2021}. As part of our study on the galaxies in the Virgo filaments, we explore the effects of the filament environment on the chemical properties of star-forming dwarf galaxies (SFDGs) in the Virgo filaments using the Sloan Digital Sky Survey (SDSS) optical spectroscopic data and Galaxy Evolution Explorer (GALEX) ultraviolet (UV) photometric data. 

This paper is structured as follows. In Section 2, we describe the selection of SFDGs in the Virgo filaments, Virgo cluster, and field. Estimations of the basic parameters of SFDGs are also presented. In Section 3, we show the results of the gas-phase metallicity and sSFR of SFDGs in the Virgo filaments in comparison with those in the Virgo cluster and field. We also present a discussion on physical processes in filaments in terms of the chemical pre-processing. Finally, we summarize our results in Section 4.  Throughout this paper, we assume a flat $\Lambda$ cold dark matter cosmology with $H_0$ = 100$h$ km s$^{-1}$ Mpc$^{-1}$, $h$ = 0.7, $\Omega_{m}$ = 0.3, and $\Omega_{\Lambda}$ = 0.7 \citep{Komatsu2011}.\\

\section{Data and Analysis}

\subsection{Selection of Dwarf Galaxies in Different Environments}

Based on the SDSS Data Release (DR) 7 and HyperLeada database, \citet{Kim2016} identified seven filaments around the Virgo cluster. The majority of galaxies associated with the Virgo filaments are low-mass dwarf galaxies ($>$ 10$^{9}$ \(M_\odot\); $\sim$87$\%$ of the total sample, \citealt{Lee2021}). We only extracted 571 galaxies in the five filaments (i.e., Virgo III, Leo Minor, Leo II A, Leo II B, and Canes Venatici) that are known to be dynamically connected to the Virgo cluster (see \citealt{Kim2016} for details).     

For an adequate comparison with galaxies in a cluster environment, we used galaxies in the Extended Virgo Cluster Catalog (EVCC; \citealt{Kim2014}). The EVCC contains a comprehensive galaxy sample (1589 galaxies) in a wide area of 725 deg$^2$ or 60.1 Mpc$^2$ reaching out to the outskirts of the Virgo cluster (i.e., 3.5 times the virial radius of the Virgo cluster).

We selected field galaxies at z $<$ 0.01 by assuming a cylinder volume constructed using the SDSS data. The circle radius of a cylinder volume was adopted to be 350 kpc that corresponds to a virial radius of an early-type galaxy with $M_{r}$ = $-$20.5 mag (\citealt{Park2008}). We adopted a radial velocity depth of $\pm$700 kms$^{-1}$, which is the 1$\sigma$ of velocity dispersion of the Virgo cluster (\citealt{Binggeli1985}). If only one galaxy was located in the cylinder volume, we defined this galaxy as a field galaxy. We secured 505 field galaxies. To obtain insights regarding the environment of the selected field galaxies, we computed their three-dimensional density contrast (i.e., over-density), $\delta_{1,1000}$, within a cylinder of a 1 h$^{-1}$ Mpc radius and 1000 kms$^{-1}$ half-length, which is defined as follows \citep{Gavazzi2010}:

\begin{equation}
   \delta_{1,1000}= \frac{\rho-\langle{\rho}\rangle} {\langle{\rho}\rangle}
\end{equation}

, where $\rho$ is the local galaxy number density, and $\langle$$\rho$$\rangle$ is the mean galaxy number density measured in the Coma/A1367 supercluster region (i.e., 0.05 galaxy (h$^{-1}$ Mpc)$^{-3}$). Most of the field galaxies have values of $\delta$ $_{1,1000}$ $<$ 0, indicating that the selected field galaxies are located in ultra-low-density environments \citep{Gavazzi2010}.

Finally, we selected dwarf galaxies in three different environments based on the morphological classification as described by \citet{Sandage1984}. The morphological classification of the EVCC was carried out by careful visual inspection of SDSS images (see \citealt{Kim2014} for details). The selected 1325 dwarf galaxies in the Virgo cluster consist of early-type dwarf galaxies, high-surface-brightness irregular galaxies, and low-surface-brightness irregular galaxies. We also selected dwarf galaxy samples in the Virgo filaments (361 galaxies) and field (336 galaxies) by conducting visual inspections of SDSS images, according to the scheme of the EVCC. Our selected dwarf galaxy samples are in the magnitude range of -19 $<$ M$_r$ $<$ -12.

\subsection{Measurement of Emission Lines}
We extracted optical spectra of our 2022 dwarf galaxies in different environments from the SDSS DR12. In the SDSS, the spectra were observed through 3$\arcsec$ fibers, giving a wavelength coverage of 3800 \AA\ to 9200 \AA\ at a spectral resolution of 1500-2500 \citep{Alam2015}. To obtain emission line fluxes, we modeled the underlying stellar continuum of the SDSS spectrum for each galaxy by using the STARLIGHT stellar population synthesis code \citep{Cid_Fernandes2004,Cid_Fernandes2005,Mateus2006}. We fitted the observed spectrum with a combination of 150 multiple simple stellar populations (6 metallicities and 25 ages) taken from the evolutionary synthetic models of \citet{Bruzual2003}. After subtracting the stellar continuum, we measured fluxes of emission lines with Gaussian profiles using the $MPFIT$ IDL routine \citep{Markwardt2009}. For correction of internal extinction to the measured emission line fluxes, we used the theoretical Balmer ratios of H$\alpha$/H$\beta$=2.86 and H$\gamma$/H$\beta$=0.47 for Case B emissivity with a temperature of 10,000 K and an electron density of 100 cm$^{-3}$ \citep{Osterbrock2006}. We only selected the galaxies with signal-to-noise ratios higher than 5 for $\oiii\lambda5007$, $\hg$, $\hb$, $\ha$, $\nii\lambda6584$, and $\sii\lambda\lambda6717,6731$ emission lines.

\subsection{Selection of Star-forming Dwarf Galaxies}

\begin{figure*}
\epsscale{1.0}
\plotone{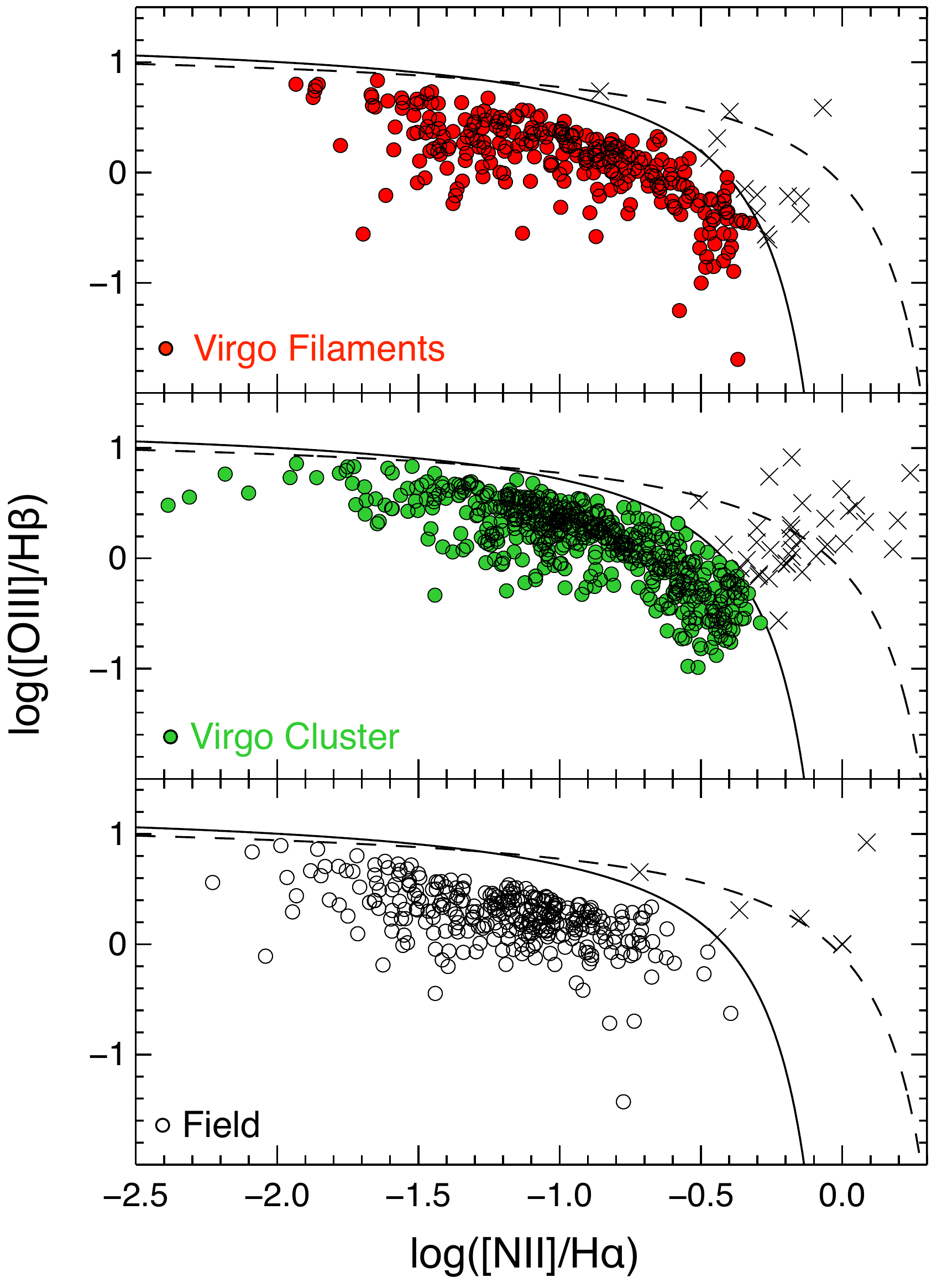}

\caption{ $\oiii\lambda5007/\hb$ vs. $\nii\lambda6584/\ha$ diagnostic diagram of dwarf galaxies in the Virgo filaments (top), Virgo cluster (middle), and field (bottom).  The circles represent the selected SFDGs that lie below the empirical star formation curve of \citet{Kauffmann2003} denoted by the solid curve. The dashed curve presents the theoretical maximum starburst model line of \citet{Kewley2001}. The crosses are AGN or composite galaxies.  }

\label{figbpt}
\end{figure*}

We identified SFDGs using the Baldwin--Phillips--Terlevich (BPT) diagram constructed from the measured emission line ratios $\oiii\lambda5007/\hb$ and $\nii\lambda6584/\ha$. The BPT is used to classify galaxies according to the excitation mechanism of their emission lines (i.e., either photoionization by young massive stars within $\hii$ regions or photoionization by non-thermal radiation from the active galactic nucleus (AGN); \citealt{Baldwin1981}). In Figure ~\ref{figbpt}, we present the BPT diagram for our sample galaxies in different environments. For the selection of SFDGs, we used the theoretical maximum starburst model line (\citealt{Kewley2001}, dashed curve) and the empirical star formation curve (\citealt{Kauffmann2003}, solid curve). We selected SFDGs that lie below the solid curve. The crosses represent composite or AGN galaxies. Finally, we secured 210, 241, and 305 SFDGs in the Virgo filaments, Virgo cluster, and field, respectively. The information of sample galaxies used in this study is summarized in Table~\ref{table1}. 

\begin{figure*}
\epsscale{1.0}
\plotone{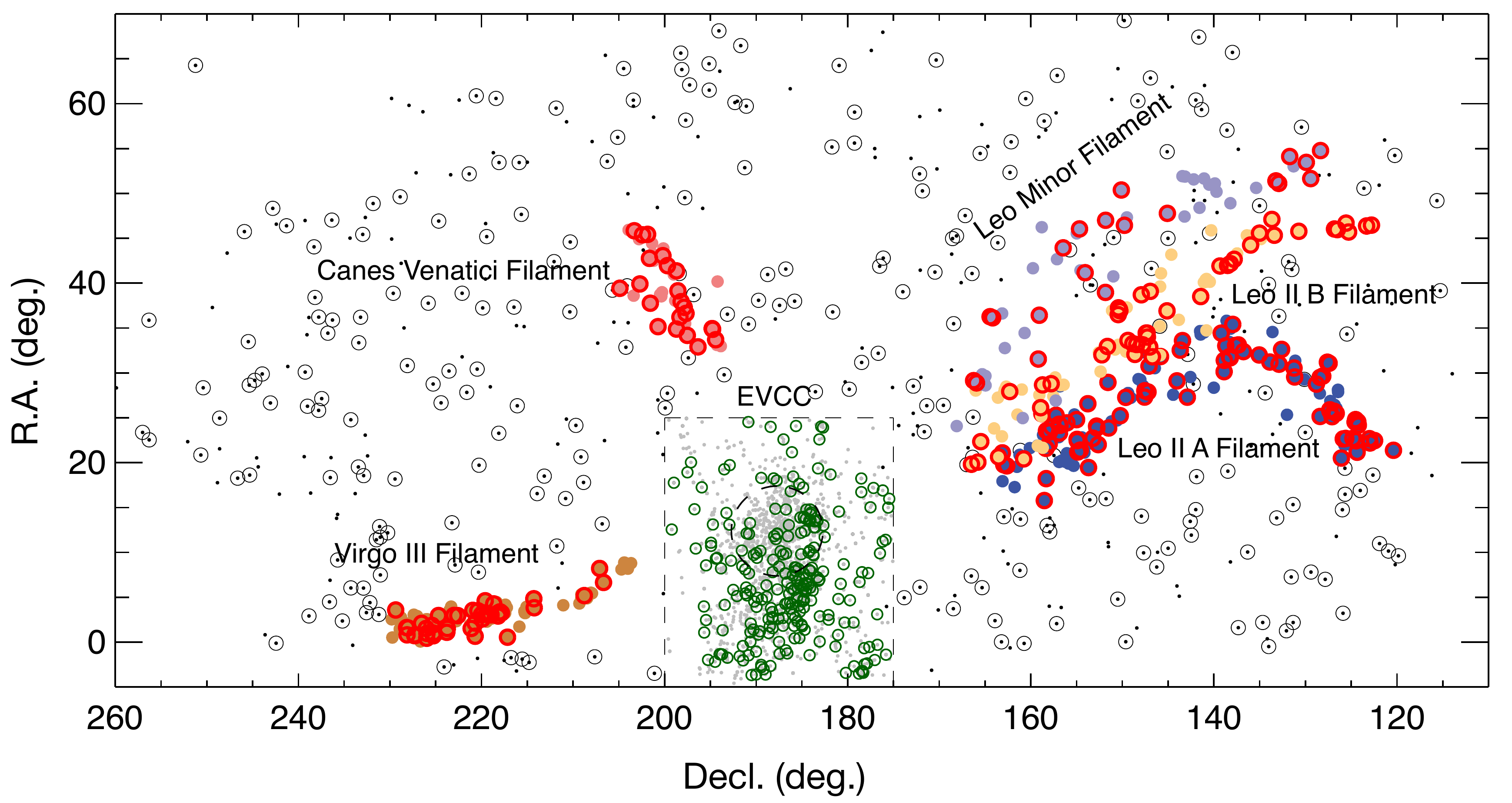} 

\caption{Projected spatial distribution of galaxies in the Virgo filaments (filled circles with different colors), Virgo cluster (gray dots within dashed rectangular box), and field (black dots). All selected SFDGs are over-plotted with open circles. The large dashed rectangular box is the region of the Extended Virgo Cluster Catalog \citep[EVCC,][] {Kim2014}. The large, dashed circle with a 6 deg radius denotes the area of one virial radius of the Virgo cluster.}   

\label{fig:spatial}
\end{figure*}

Figure~\ref{fig:spatial} shows the sky distribution of galaxies in the Virgo filaments (filled circles with different colors), Virgo cluster (gray dots within the EVCC area), and field (black dots). The selected SFDGs are over-plotted with open circles.

\subsection{Morphological Subsamples of Star-forming Dwarf Galaxies}

\begin{figure*}
\epsscale{1.1}
\plotone{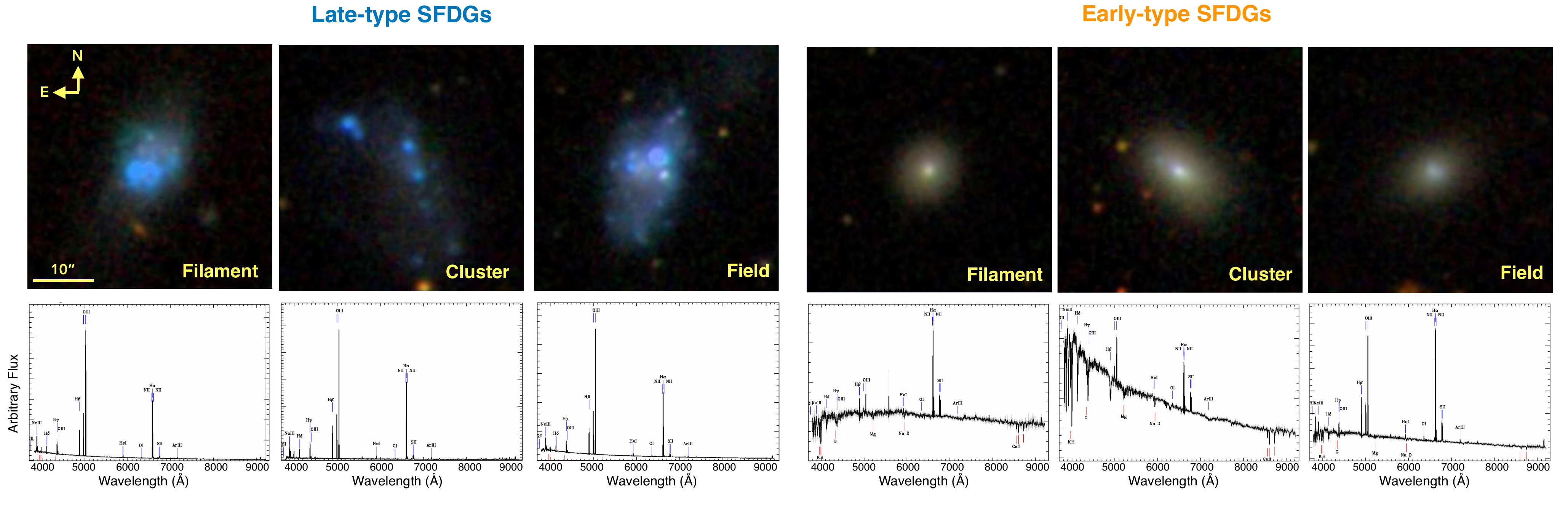} 

\caption{Top: Example SDSS $g$, $r$, and $i$ composite images of L-SFDGs (left three panels) and E-SFDGs (right three panels) in different environments. Bottom: SDSS optical spectra of galaxies at their centers within a 3$\arcsec$ fiber diameter, displaying the emission lines}

\label{fig:morp}
\end{figure*}

We divided SFDGs into two subsamples based on their morphologies: early-type SFDGs (E-SFDGs) and late-type SFDGs (L-SFDGs). The E-SFDGs and L-SFDGs show morphologies of early-type dwarf galaxies and dwarf irregular galaxies, respectively. The E-SFDGs are characterized by spheroidal shapes and smooth, featureless envelopes. The outer parts of E-SFDGs are generally red, but their central regions show blue colors. The L-SFDGs present irregular envelopes with overall blue colors and are dominated by single or multiple HII regions. Some L-SFDGs exhibit elliptical shapes in the outer part, but they show overall blue colors and irregular structures at their centers. In Figure ~\ref{fig:morp}, we present examples of SDSS color images of E-SFDGs and L-SFDGs in the Virgo filaments, Virgo cluster, and field.  

The morphological characteristics of E-SFDGs and L-SFDGs are also consistent with their optical spectra (see the bottom panels of Figure ~\ref{fig:morp}).
The L-SFDGs show strong narrow emission lines, indicative of intense star formation. Emission lines such as \ha$, \hb$, and $\oiii$ are
predominant compared to the continuum in optical spectra (see the bottom left three panels of Figure 3). However, E-SFDGs have relatively weak emission lines and are less dominant than the continuum level (see the bottom right three panels of Figure ~\ref{fig:morp}). However, optical spectra reveal that E-SFDGs are objects with hints of recent or ongoing star formation at their central regions.

\subsection{Gas-phase Metallicity, Star Formation Rate, and Stellar Mass}

The gas-phase metallicity of a galaxy is usually specified by the oxygen abundance, the most abundant metal within its interstellar medium. We estimated gas-phase oxygen metallicity using the following empirical relationship derived by \citet{Denicolo2002}:  

\begin{equation}
  12+log(O/H)=9.12+N2\times0.73
\end{equation}

, where N2 is the metallicity indicator of N2 = log($\nii\lambda6584/\ha$) \citep{Denicolo2002}.

Total UV flux from the GALEX mission reflects the overall star formation of a galaxy. GALEX provides UV sky surveys with wide area coverage and deep sensitivity in two UV bands, far-UV (FUV; 1350-1750\AA) and near-UV (NUV; 1750-2750\AA). The UV fluxes of SFDGs were taken from the GALEX archive. We derived the NUV star formation rates (SFRs) of SFDGs from the relation given by \citet{Kennicutt1998} based on the NUV luminosity ($L_{NUV}$):
\begin{equation} 
SFR_{NUV}(M_\odot yr^{-1})=1.4\times10^{-28}L_{NUV}(erg \, s^{-1} Hz^{-1}).
 \end{equation} \

The stellar masses of SFDGs were derived using the relation between the SDSS $g-i$ color and the stellar mass-to-light ratio based on the $i$-band luminosity ($L_{i}$; \citealt{Bell2003}) assuming the initial mass function of \citet{Kroupa1993}:

\begin{equation} 
log(M_\star/M_\odot)=-0.152+0.518(g-i)+log(L_i/L_\odot). 
 \end{equation} 

For the SFR and stellar mass calculations, based on the reddening law of \citet{Cardelli1989}, we corrected galactic extinctions for the magnitudes using the E($B-V$) value from the NASA/IPAC Extragalactic Database. The distances of SFDGs in the Virgo filaments and field were estimated using their radial velocities. In the case of the Virgo cluster, we assumed the distances of all SFDGs to be 17 Mpc by adopting the typical distance of the main body of the Virgo cluster \citep{Boselli2014}.

\section{Results and Discussion}

%%%%%%%%%%%%%%%%%%%%%%%%%%%%%%%%%%%%%%%%%%%%%%%%%%%%
%%%%%%%%%%%%%%%%%%%%%%%%%%%%%%%%%%%%%%%%%%%%%%%%%%%%
% Stellar Mass--O/H and Stellar Mass--sSFR Relations
%%%%%%%%%%%%%%%%%%%%%%%%%%%%%%%%%%%%%%%%%%%%%%%%%%%%
%%%%%%%%%%%%%%%%%%%%%%%%%%%%%%%%%%%%%%%%%%%%%%%%%%%%

\subsection{O/H vs. Stellar Mass and Specific Star Formation Rate vs. Stellar Mass Distributions}  

%%%%%%%%%%%%%%%%%%%%
% Filament vs. Cluster
%%%%%%%%%%%%%%%%%%%%

\begin{figure*}[t]
\epsscale{1.0}
\plotone{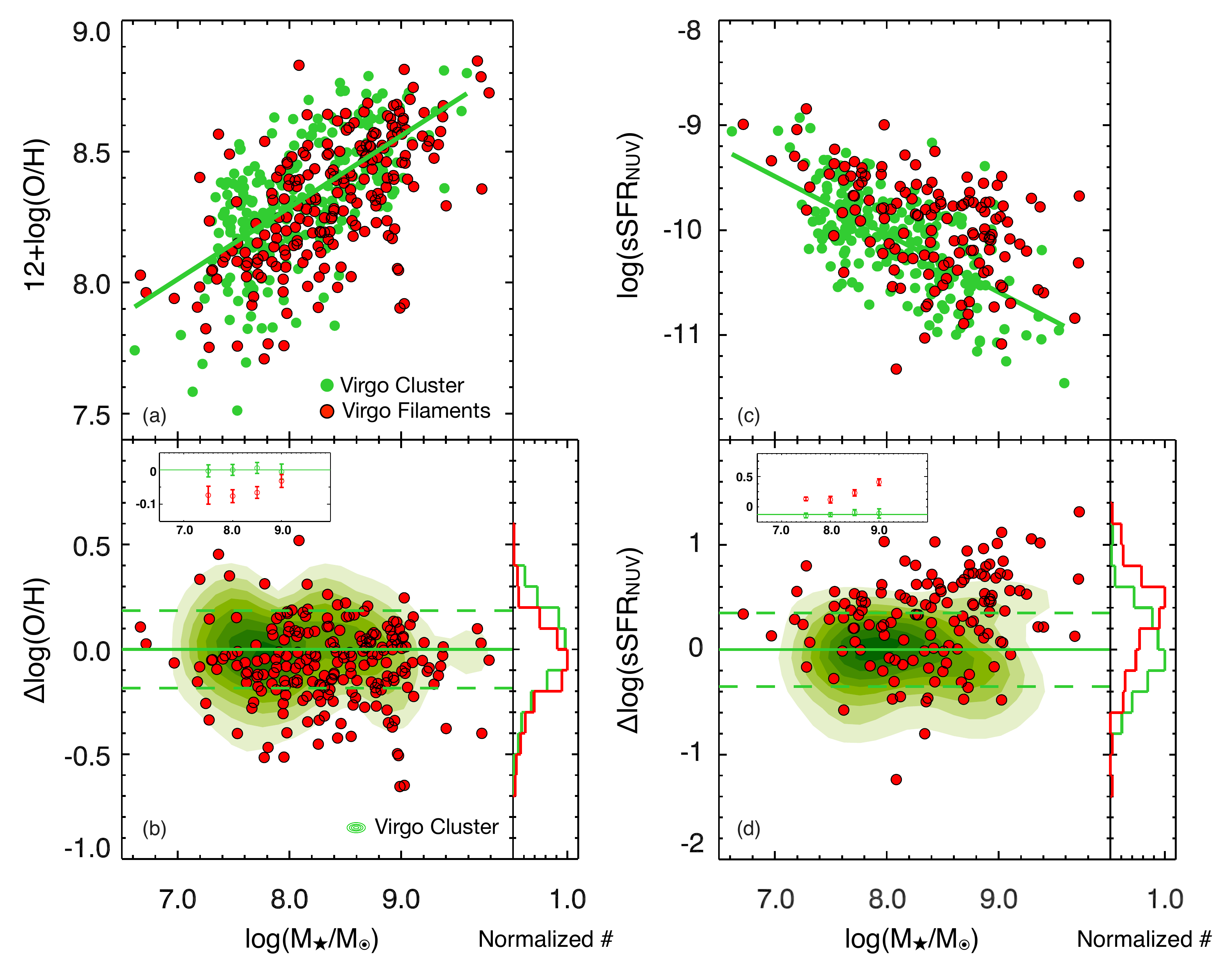} 
\caption{$(a)$ and $(c)$: O/H vs. stellar mass and sSFR vs. stellar mass distributions of SFDGs between the Virgo filaments (red filled circles) and Virgo cluster (green filled circles). The green solid line denotes the best linear fit to the SFDGs in the Virgo cluster. $(b)$ and $(d)$: The residual distributions of the O/H and sSFR from the linear fit to the Virgo cluster. The contours are the distributions of SFDGs in the Virgo cluster. Dashed lines are 1$\sigma$ deviations from the mean of SFDGs in the Virgo cluster. Insets represent residuals of SFDGs in the Virgo cluster (green open circles) and Virgo filaments (red open circles) at different mass bins. The error bars are uncertainties obtained from 10,000 bootstrap resampling. Green and red histograms represent residual distributions of SFDGs in the Virgo cluster and Virgo filaments, respectively.}

\label{fig:F_C}
\end{figure*}

In Figure ~\ref{fig:F_C}(a) and (c), we present the distributions of O/H vs. stellar mass and sSFR vs. stellar mass for SFDGs in the Virgo filaments (red filled circles) and Virgo cluster (green filled circles). The error-weighted best linear fit to the SFDGs in the Virgo cluster is shown as the green solid line: 12+log(O/H)=0.27log(M$_\star$/M$_\odot$)+6.10 for O/H vs. stellar mass and log(sSFR)=$-$0.55log(M$_\star$/M$_\odot$)$-$5.62 for sSFR vs. stellar mass. The O/H and sSFR residuals from the linear fit to the Virgo cluster are also presented in Figure ~\ref{fig:F_C}(b) and (d). In these plots, the distributions of SFDGs in the Virgo cluster are shown as contours. Dashed lines are 1$\sigma$ deviations from the mean of SFDGs in the Virgo cluster. Insets represent residuals and errors of SFDGs at different mass bins. Green and red histograms represent residual distributions of SFDGs in the Virgo filaments and Virgo cluster, respectively.

While there is obvious scatter, the SFDGs in the Virgo filaments and Virgo cluster appear to show different distributions in Figure ~\ref{fig:F_C}. The SFDGs in the Virgo filaments deviate toward a lower O/H than those in the Virgo cluster at a given stellar mass. However, SFDGs in the Virgo filaments show a higher mean sSFR at a given stellar mass compared to their Virgo cluster counterparts, which is consistent with the well-known anti-correlation between O/H and sSFR \citep{LaraLopez2010,Mannucci2010,Hunt2012,Yates2012}. The Kolmogorov–Smirnov(K-S) tests for O/H and sSFR residuals between SFDGs in the Virgo filaments and Virgo cluster yielded probabilities of $P$ $\approx$ 0.007 and $P$ $\approx$ 0, respectively. These tests indicate that the differences in O/H and sSFR distributions between SFDGs in the Virgo filaments and Virgo cluster are statistically significant.

%%%%%%%%%%%%%%%%%%%%
% Filament vs. field
%%%%%%%%%%%%%%%%%%%%

\begin{figure*}[t]
\epsscale{1.0}
\plotone{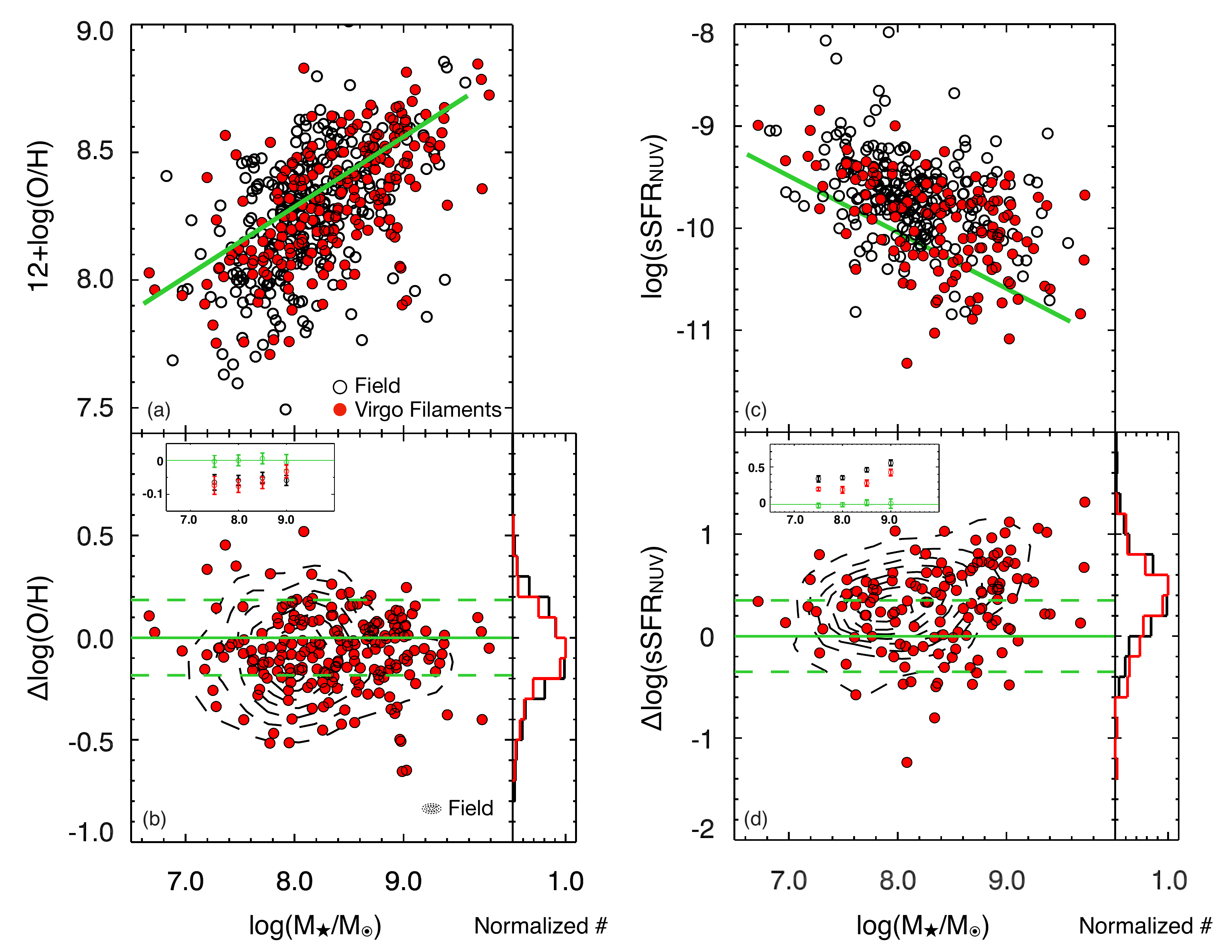} 

\caption{ $(a)$ and $(c)$: O/H vs. stellar mass and sSFR vs. stellar mass distributions of SFDGs between the Virgo filaments (red filled circles) and field (black circles). The green solid line denotes the best linear fit to the SFDGs in the Virgo cluster. $(b)$ and $(d)$: The residual distributions of the O/H and sSFR from the linear fit to the Virgo cluster. The contours are the distributions of SFDGs in the field. Dashed lines are 1$\sigma$ deviations from the mean of SFDGs in the Virgo cluster. Insets represent residuals of SFDGs in the Virgo cluster (green open circles), Virgo filaments (red open circles), and field (black open circles) at different mass bins. The error bars are uncertainties obtained from 10,000 bootstrap resampling. Black and red histograms represent residual distributions of SFDGs in the field and Virgo filaments, respectively.}

\label{fig:F_F}
\end{figure*}

    In Figure~\ref{fig:F_F}, we present the distributions of O/H vs. stellar mass and sSFR vs. stellar mass for SFDGs in the Virgo filaments (red filled circles) and field (black open circles). For comparison, the best linear fit to the SFDGs in the Virgo cluster is shown as the green solid line. The O/H and sSFR residuals from the linear fit to the Virgo cluster are also presented in Figure ~\ref{fig:F_F}(b) and (d). In these plots, the distributions of SFDGs in the field are shown as contours. Dashed lines are 1$\sigma$ deviations from the mean of SFDGs in the Virgo cluster. Insets represent residuals and errors of SFDGs at different mass bins. Black and red histograms represent residual distributions of SFDGs in the field and Virgo filaments, respectively.

As shown in Figure~\ref{fig:F_F}, the SFDGs in the Virgo filaments tend to share similar distributions with those in the field without showing a significant systematic difference.  The K-S tests for O/H and sSFR residual distributions between SFDGs in the Virgo filaments and field yielded probabilities of $P$ $\approx$ 0.81 and $P$ $\approx$ 0.78, respectively. This indicates that the O/H and sSFR distributions between SFDGs in the Virgo filaments and field are not statistically different.

%%%%%%%%%%%%%%%%%%%%%%%%%%%%%%%%%%%%%%%%%%%%%%%%%%%%
%%%%%%%%%%%%%%%%%%%%%%%%%%%%%%%%%%%%%%%%%%%%%%%%%%%%
% Effects of the Environment
%%%%%%%%%%%%%%%%%%%%%%%%%%%%%%%%%%%%%%%%%%%%%%%%%%%%
%%%%%%%%%%%%%%%%%%%%%%%%%%%%%%%%%%%%%%%%%%%%%%%%%%%%

\subsection{Effects of the Environment}  

There has been debate about the role of  environmental effects on the metallicity and star formation activity in star-forming galaxies \citep{Skillman1996,Hughes2013}. An environmental process related to the hot intergalactic medium, such as ram pressure stripping, is vital for chemical enrichment and star formation regulation in galaxy clusters \citep{Gunn1972}. 
If the ram pressure exceeds the gravitational restoring force of a galaxy, a significant fraction of the interstellar medium in the outskirts of the galaxy would be removed. This process can cause a deficit in the pristine gas reservoir of the galaxy.

Ram pressure stripping is primarily dominant in the central region of a cluster \citep[e.g., inside of one virial radius of the cluster;][]{Tonnesen2007}. In the case of low-mass star-forming galaxies that reside in the cluster core, ram pressure stripping can efficiently remove the HI interstellar gas, which gives rise to chemical enrichment and the suppression of star formation \citep[e.g.,][]{Petropoulou2012}. However, ram pressure in the outskirts of the clusters outside of one virial radius is expected to be $\sim$100 times lower than that in a core of the cluster \citep{Tecce2011}. In the case of the Virgo cluster, the majority ($\sim$78$\%$; 187/241) of SFDGs reside beyond the region of one virial radius from the cluster center (R$_{vir}$ = 5$\degr$.21 or 1.55 Mpc; \citealt{Ferrarese2012}; see also the dashed circle in Figure ~\ref{fig:spatial}), indicating that most
SFDGs in the Virgo cluster would be less sensitive to the ram pressure stripping effect. 

\begin{figure*}
\epsscale{1.0}
\plotone{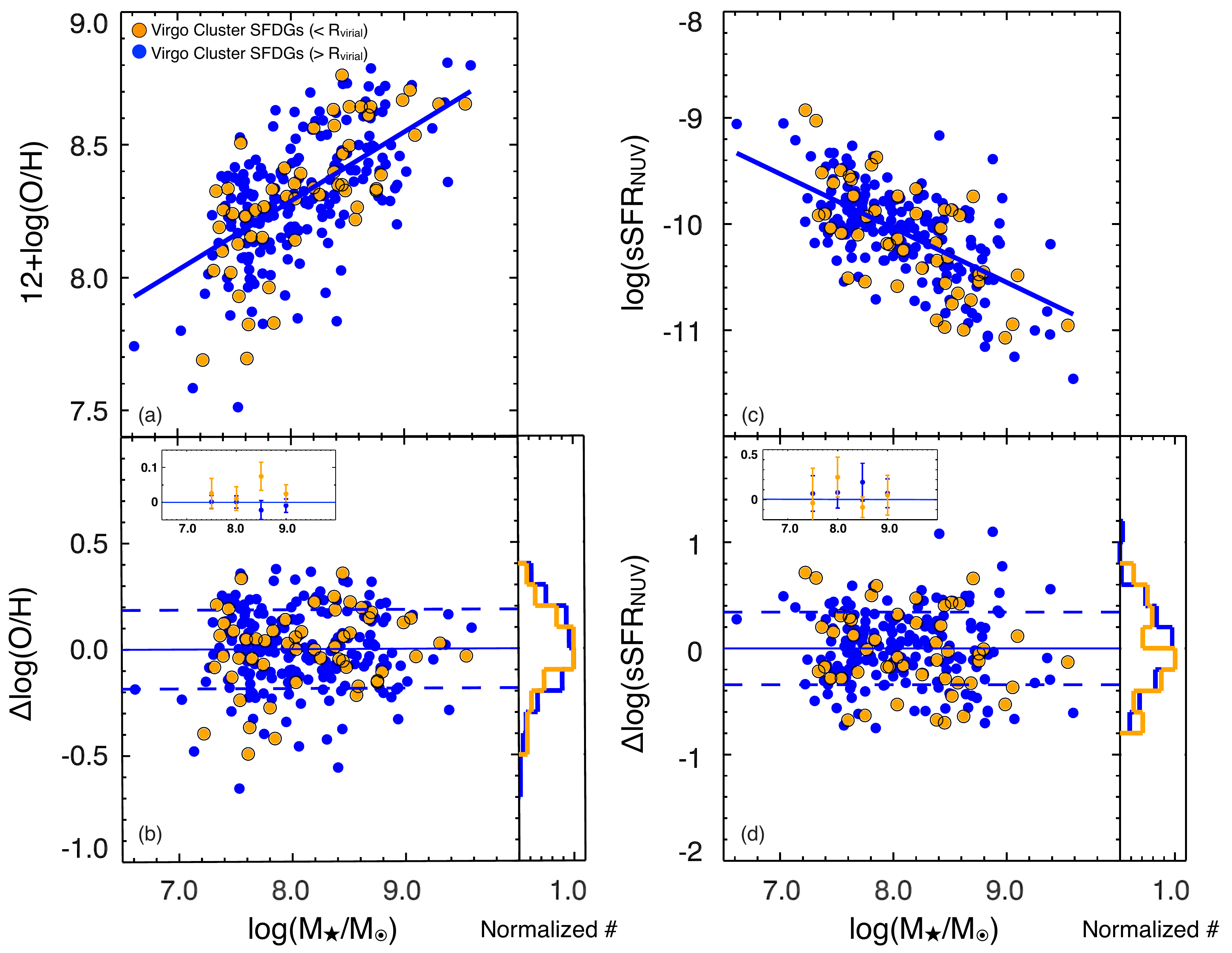} 
\caption{$(a)$ and $(c)$: O/H vs. stellar mass and sSFR vs. stellar mass distributions of two SFDG subsamples located inside (yellow filled circles) and outside (blue filled circles) of one virial radius of the Virgo cluster. The blue solid line denotes the best linear fit to the SFDGs located outside of one virial radius of the Virgo cluster. $(b)$ and $(d)$: The residual distributions of the O/H and sSFR from the linear fit to the SFDGs located outside of one virial radius of the Virgo cluster. Dashed lines are 1$\sigma$ deviations from the mean. Insets represent residuals of two subsamples. The error bars are uncertainties obtained from 10,000 bootstrap resampling. Histograms represent residual distributions of two subsamples.}

\label{fig:insideout}
\end{figure*}

By dividing SFDGs in the Virgo cluster into two subsamples located inside (yellow filled circles) and outside (blue filled circles) of one virial radius of the Virgo cluster, we present their O/H vs. stellar mass and sSFR vs. stellar mass distributions in Figure 6. In Figure~\ref{fig:insideout}(a) and (c), the error-weighted best linear fit to the SFDGs located outside of one virial radius of the Virgo cluster is shown as the blue solid line. The O/H and sSFR residuals from this linear fit are also presented in Figure ~\ref{fig:insideout}(b) and (d). If SFDGs in the cluster core suffer more gas stripping by ram pressure than their counterparts in the cluster outskirts, the distributions of O/H and sSFR between two subsamples are expected to differ. However, as shown in Figure~\ref{fig:insideout}, the two subsamples show no distinct different distributions in both panels.
The K-S tests for O/H and sSFR residuals between two subsamples yielded probabilities of P$\sim$0.97 and P$\sim$0.69, respectively. This indicates that the O/H and sSFR distributions between SFDGs located inside and outside of one virial radius of the Virgo cluster are not statistically different. We suggest that ram pressure is not the primary driver of the metallicity enhancement and star formation suppression in SFDGs of the Virgo cluster.

The interaction between galaxies is considered as one of the typical environmental processes. The tidal forces in galaxy interactions can remove a large fraction of gas from galaxies, which leads to a subsequent decrease in star formation activity and the enhancement of metallicity \citep{Barnes1991,Mihos1994,Mihos1996,DiMatteo2007,Ellison2009}.
\cite{Ellison2009} have shown that gas-phase metallicity is mainly affected by the local environment in which galaxies with a close companion galaxy in locally higher-density environments show higher O/H than those with no near neighbors. 

To quantify the effectiveness of interactions between galaxies, we calculated the perturbation parameter $f$ of SFDGs in the Virgo filaments and Virgo cluster. The $f$ value is defined by \citet{Varela2004} as 

\begin{equation}
    f=log(\frac {F_{ext}} {F_{int}})=3log(\frac{R} {D_{P}})+0.4(m_{G}-m_{P})
\end{equation}

, where F$_{ext}$ indicates the tidal force exerted by the primary galaxy and F$_{int}$ is the internal force in the outskirts of the secondary galaxy; $R$ is the size of the secondary galaxy; D$_{p}$ is the projected distance between the primary and the secondary; and m$_{G}$ and m$_{p}$ are the apparent magnitudes of the secondary and the primary, respectively. The $f$ value signifies the ratio between the internal force and external tidal force acting on the galaxy by a possible perturber. For the $f$ value calculation of each SFDG, we considered all SDSS galaxies with a relative velocity of less than 500 kms$^{-1}$ within its environment. The galaxy with the highest $f$ value is chosen to be the perturber that exerts the most significant tidal force on the SFDG. We defined galaxies with $f$ $>$ $-$4.5 as those under the influence of tidal perturbation  \citep{Varela2004}. The $f$ value can be considered as a proxy of the local environment since this value depends on the distance and luminosity ratio between two galaxies. 

\begin{figure*}
\epsscale{1.0}
\plotone{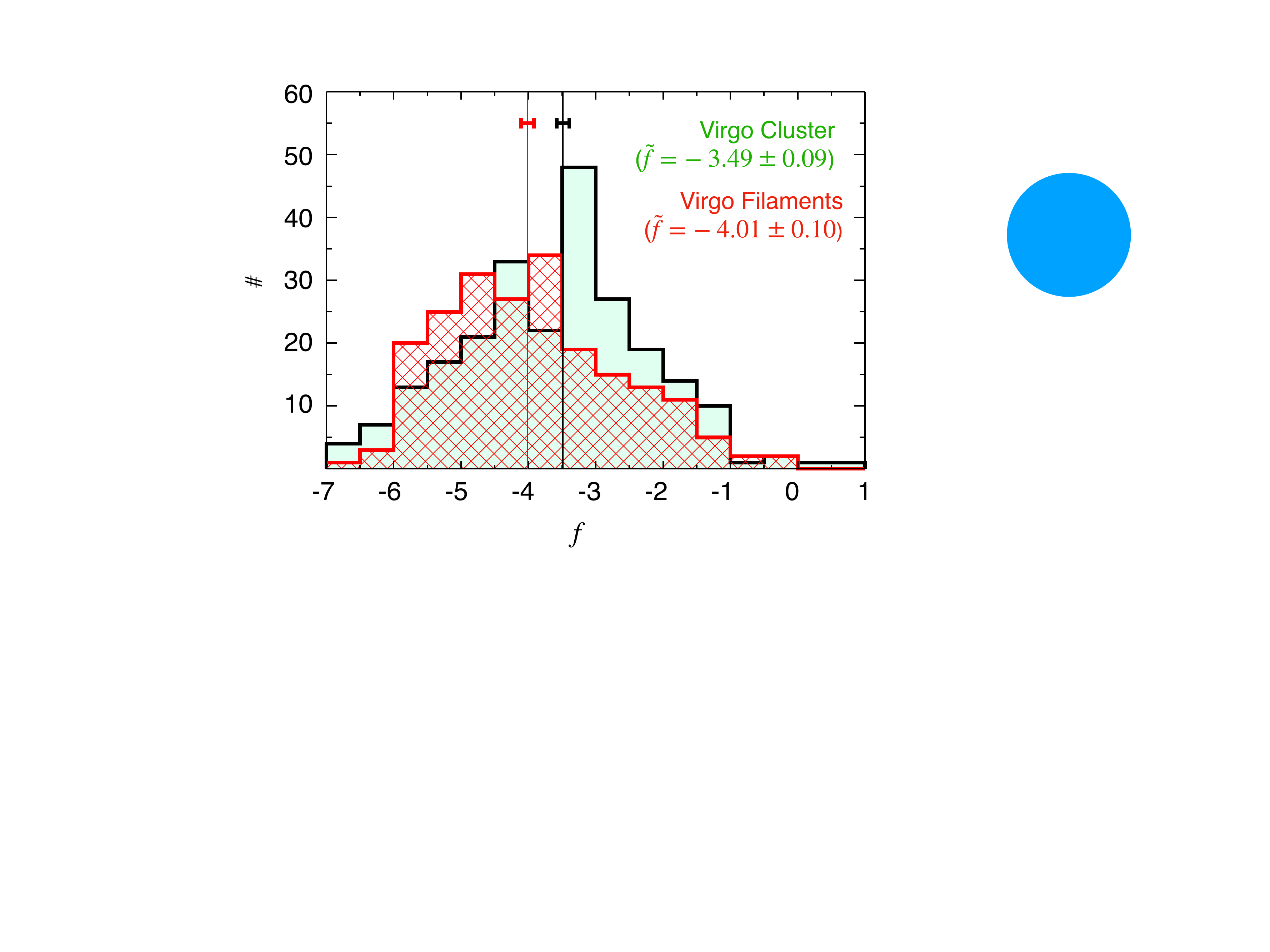} 
\caption{Distributions of $f$ values of SFDGs in the Virgo filaments (red histogram) and Virgo cluster (light green histogram). The red and black vertical lines denote the median $f$ values of SFDGs in the Virgo filaments and Virgo cluster, respectively. The error bars are uncertainties obtained from 10,000 bootstrap resampling. The median and error of $f$ values of SFDGs in the Virgo filaments and Virgo cluster are also given.} 

\label{fig:fv_total}
\end{figure*}

Figure~\ref{fig:fv_total} shows the $f$ value distributions of SFDGs in the Virgo filaments (red histogram) and Virgo cluster (light green histogram). The red and black vertical lines represent the median $f$ values for the SFDGs in the Virgo filaments and Virgo cluster, respectively. The error bars are uncertainties obtained from 10,000 bootstrap resampling. The most notable feature is that the $f$ value distribution of SFDGs in the Virgo filaments appears different from those in the Virgo cluster. The SFDGs in the Virgo filaments show a skewed distribution toward lower $f$ values (with a median value of $-$4.01$\pm$0.10), whereas the distribution of SFDGs in the Virgo cluster has a higher peak (with a median value of $-$3.49$\pm$0.09). The K-S test for $f$ value distributions between SFDGs in the Virgo filaments and Virgo cluster yielded a probability of $P$ $\approx$ 0, indicating a statistically significant difference between the two distributions. Consequently, this suggests that tidal interactions between galaxies are likely to prevail in the Virgo cluster compared to the Virgo filaments, which would be responsible for the chemically evolved SFDGs with suppressed star formation activity in the Virgo cluster.

%-----------------------------------------------------------------------------------------------------------------------------------------

\subsection{Environmental Disparity between Virgo filaments}

\begin{figure*}
\epsscale{1.0}
\plotone{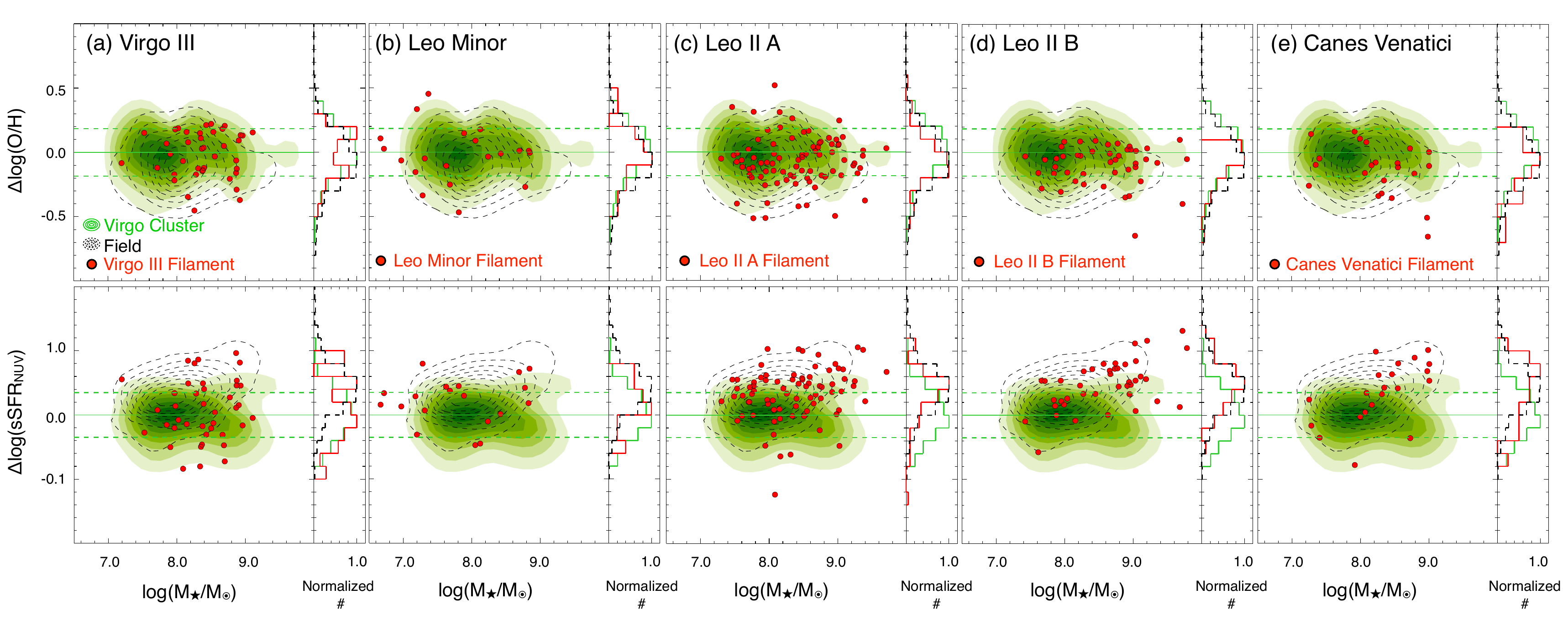} 
\caption{Residual distributions of the O/H (upper panels) and sSFR (lower panels) for SFDGs (red filled circles) in each filament obtained from the best linear fit of O/H vs. stellar mass and sSFR vs. stellar mass distributions for the Virgo cluster (green solid line). Dashed lines are 1$\sigma$ deviations from the mean of SFDGs in the Virgo cluster. The symbols and contours are the same as those in Figures ~\ref{fig:F_C} and ~\ref{fig:F_F}.}

\label{fig:fig_residuals}
\end{figure*}

We examined the properties of SFDGs in individual Virgo filament by comparing them with those in the Virgo cluster and field. Figure~\ref{fig:fig_residuals} shows the O/H (upper panels) and sSFR (lower panels) residuals from the linear fit to the Virgo cluster for each Virgo filament. Symbols are the same as those in Figures ~\ref{fig:F_C} and ~\ref{fig:F_F}. Although there are small number statistics and scatter, the residual O/H distribution of the Virgo III filament appears different from those of the other four filaments. While SFDGs in the Virgo III filament are located around the mean of the Virgo cluster (green solid line), SFDGs in the other filaments are slightly (e.g., Leo Minor and Leo II A) or distinctly (Leo II B and Canes Venatici) deviated toward the negative residuals. Moreover, SFDGs in the Virgo III filament show a similar distribution to that of the Virgo cluster (green contours), whereas SFDGs in the other filaments exhibit rather field-like distributions (dashed contours). The difference between the Virgo III and other filaments is more significant in the sSFR residual distributions. About half of all SFDGs in the Virgo III filament have negative residuals (i.e., lower sSFRs than the mean of the Virgo cluster) that overlap with the distribution of the Virgo cluster. However, most SFDGs in the other filaments show systematically positive residuals (i.e., higher sSFRs than the mean of the Virgo cluster) consistent with the distribution of field SFDGs. The similarity of O/H and sSFR of SFDGs in the Virgo III filament with those in the Virgo cluster implies that the properties of SFDGs in the Virgo III filament may have already been changed before they fell into the Virgo cluster (i.e., 'pre-processing').

Several environmental effects are suggested as possible mechanisms for the galaxy pre-processing in filaments. Galaxies moving along the filaments can accrete cold gas from the intrafilament medium that replenishes some HI gas into filament galaxies \citep{Keres2005,Sancisi2008,Darvish2015,Kleiner2017}. \citet{Darvish2015} found that star-forming galaxies in filaments at z$\sim$ 0.5, on average,  have higher metallicity than those in the field, probably owing to the inflow of the pre-enriched intrafilamentary gas into filament galaxies. They also showed that the electron densities of the filament galaxies are significantly lower than those of the field counterparts, which might be due to a longer star formation timescale by gas accretion for filament star-forming galaxies.

We calculated the electron densities of SFDGs in the Virgo III filament in comparison to the field SFDGs using the $\sii\lambda6717/\lambda6731$ ratios \citep{Osterbrock2006}. The median $\sii\lambda6717/\lambda6731$ ratios with 1$\sigma$ errors of SFDGs in the Virgo III filament and field are 1.33$\pm$0.13 and 1.32$\pm$0.17, respectively. The K-S test for electron density distributions yielded a probability of $P$ $\approx$ 0.55, indicating no significant difference in the electron densities of SFDGs between the Virgo III filament and field. Moreover, it is worth noting that gas accretion is only enabled for the massive galaxies in filaments (e.g., $>$10$^{11}$\solmass; \citealt{Kleiner2017}) due to their large gravitational potentials that are sufficient to pull cold gas from the intrafilament medium. Consequently, we suggest that the inflow of pre-enriched intrafilamentary gas is not responsible for the observed chemical enrichment of SFDGs, that are in a mass range of $<$10$^{10}$\solmass, in the Virgo III filament .    

\begin{figure*}
\epsscale{0.7}
\plotone{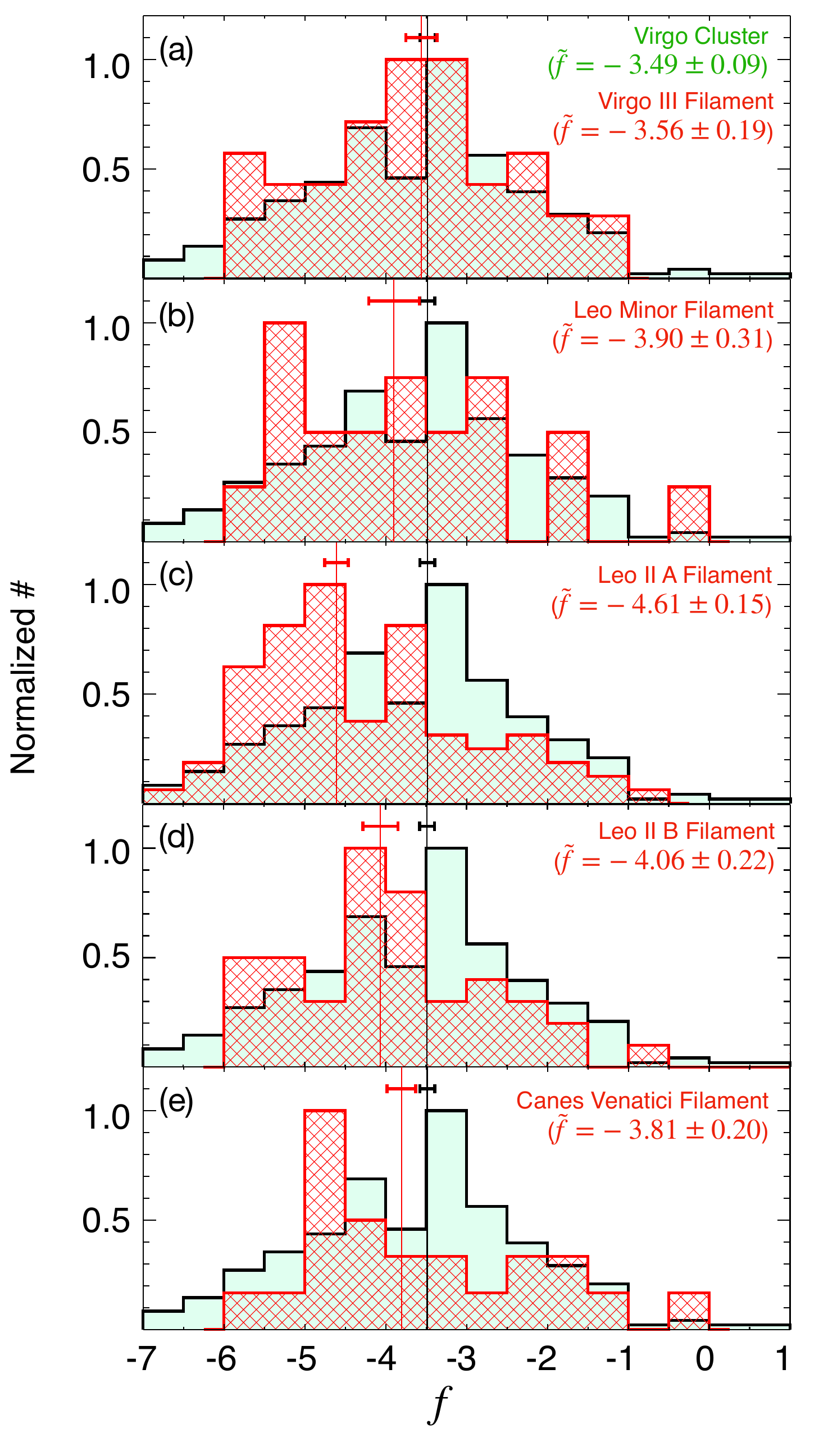} 
\caption{Distributions of $f$ values of SFDGs in each filament (red histogram) and Virgo cluster (light green histogram). The red and black vertical lines denote the median $f$ values of SFDGs in the Virgo filaments and Virgo cluster, respectively. The error bars are uncertainties obtained from 10,000 bootstrap resampling. The median and error of $f$ values of SFDGs in each filament are also given. }

\label{fig:fv}
\end{figure*}

We calculated the perturbation parameter $f$ values of SFDGs in each Virgo filament, which trace the effectiveness of galaxy interactions. Figure ~\ref{fig:fv} shows the $f$ value distributions of SFDGs in each filament (red histogram) by comparing them with that of the Virgo cluster (light green histogram). The red and black vertical lines represent the median $f$ values for each filament and the Virgo cluster, respectively. The error bars are uncertainties obtained from 10,000 bootstrap resampling.  The most interesting feature in Figure~\ref{fig:fv}(a) is that the distribution of the Virgo III filament (with a median value of $-3.56$$\pm$0.19) is remarkably similar to that of the Virgo cluster (with a median value of -$3.49$$\pm$0.09). A large fraction of SFDGs in the Virgo III filament shows high $f$ values with $>$ $-$4.5 that are considered as objects under the influence of tidal perturbation  \citep{Varela2004}. However, SFDGs in the other four filaments show different distributions from that of the Virgo cluster in which most SFDGs deviate to the lower $f$ values (see also the lower median values given in Figure~\ref{fig:fv}(b)-(e)). This result suggests that SFDGs in the Virgo III filament are in a different local environment from those in the other filaments, but are rather comparable to the Virgo cluster where tidal interactions between galaxies are more frequent.

\subsection{Transitional Dwarf Galaxies}

In the context of the morphologies of galaxies related with the physical conditions of the environment in which galaxies are located \citep{Dressler1980,Binggeli1987}, we examined the morphological properties of SFDGs in the Virgo filaments. The L-SFDGs with overall irregular shapes and blue colors are the dominant population (82\%, 172/210) in the Virgo filaments. However, it is interesting to note that a fraction of E-SFDGs (18\%, 38/210), which are characterized by overall elliptical shapes with red colors but bluer colors at their centers, are also found (see Section 2.4 and Figure~\ref{fig:morp}).

\begin{figure*}
\epsscale{1.0}
\plotone{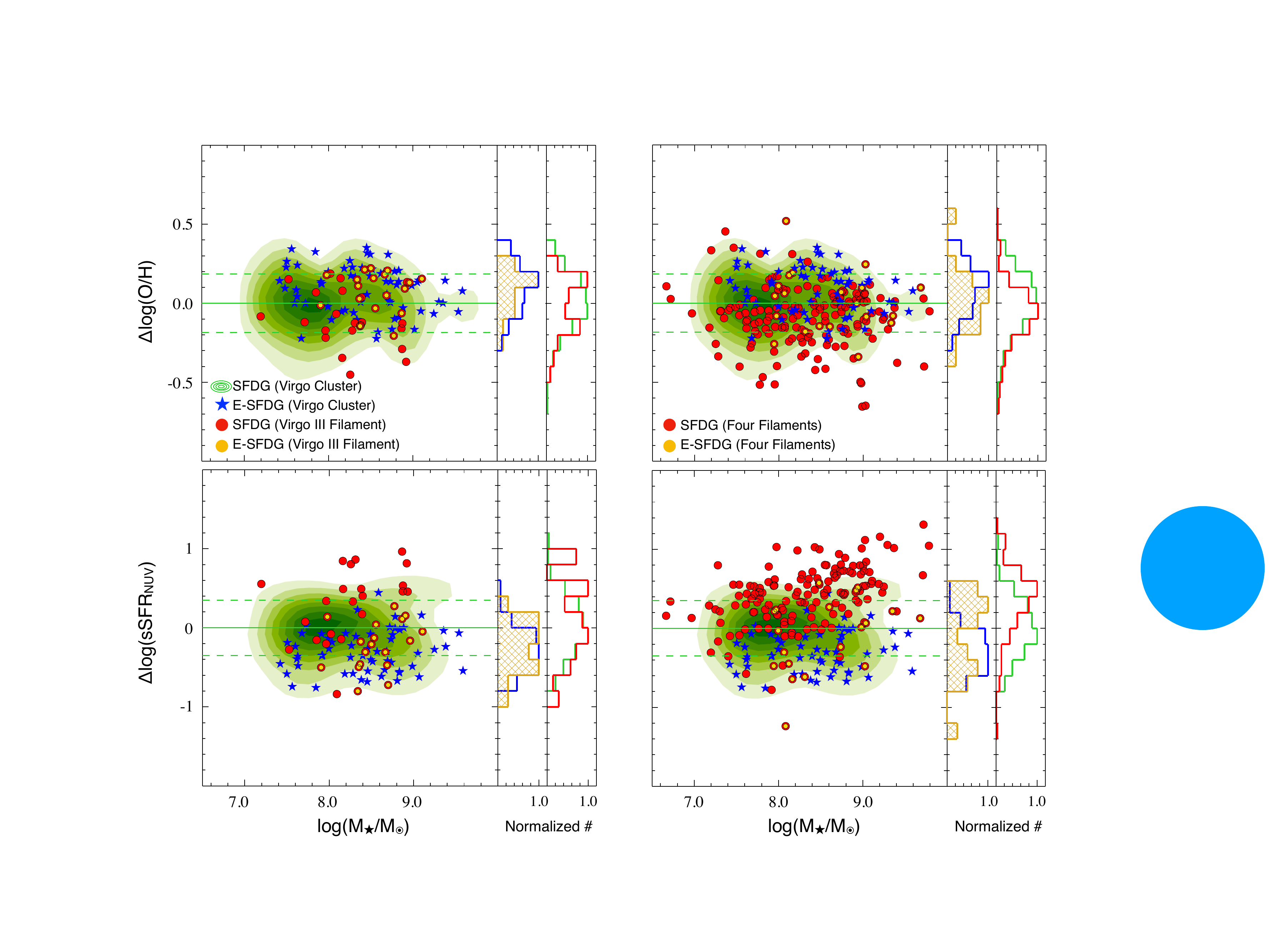} 

\caption{Residual distributions of the O/H (upper panels) and sSFR (lower panels) for SFDGs in the Virgo III filament (left panels) and other four filaments (right panels) obtained from the best linear fit of O/H vs. stellar mass and sSFR vs. stellar mass distributions of the Virgo cluster (green solid line). Dashed lines are 1$\sigma$ deviations from the mean of SFDGs in the Virgo cluster. In each panel, small yellow and large red circles denote the E-SFDGs and all SFDGs in the Virgo filaments, respectively. Blue stars and green contours are E-SFDGs and all SFDGs in the Virgo cluster. Histograms represent distributions of different subsamples of SFDGs; E-SFDGs in the Virgo filaments (brown), E-SFDGs in the Virgo cluster (blue), all SFDGs in the Virgo filaments (red), and all SFDGs in the Virgo cluster (green). }

\label{fig:E-SFDG}
\end{figure*}

In Figure~\ref{fig:E-SFDG}, we present the residual O/H and sSFR distributions of E-SFDGs (small yellow circles) and all SFDGs (large red circles) in the Virgo III filament and the other four filaments by comparing them with E-SFDGs (blue stars) and all SFDGs (green contours) in the Virgo cluster. Histograms also reflect the O/H and sSFR distributions of different subsamples of SFDGs; E-SFDGs in the Virgo filaments (brown), E-SFDGs in the Virgo cluster (blue), all SFDGs in the Virgo filaments (red), and all SFDGs in the Virgo cluster (green). Overall, the histograms representing E-SFDGs show skewed distributions toward the higher metallicities and lower sSFRs compared to those of all SFDGs regardless of the environment. This result implies that the E-SFDGs are chemically evolved objects with low star formation activities.

One interesting feature is that both the distributions of the O/H and sSFR of E-SFDGs in the Virgo III filament are very similar to those of the Virgo cluster. The median values and bootstrap resampling errors of [$\Delta$log(O/H), $\Delta$log(sSFR)] are [0.13 $\pm$ 0.03, $-$0.18 $\pm$ 0.07] and [0.12 $\pm$ 0.02, $-$0.30 $\pm$ 0.03] for E-SFDGs in the Virgo III filament and Virgo cluster, respectively. However, E-SFDGs in the other four filaments show a hint of slightly different distributions from those in the Virgo cluster. The E-SFDGs in the other filaments appear to be, on average, more metal-poor and have higher sSFR than their counterparts in the Virgo cluster. The median values and bootstrap resampling errors of [$\Delta$log(O/H), $\Delta$log(sSFR)] are [0.05 $\pm$ 0.04, 0.01 $\pm$ 0.10] for E-SFDGS in the four filaments. Moreover, it is worth noting that a substantial fraction of SFDGs ($\sim$45$\%$, 18/40) are E-SFDGs in the Virgo III filament, while only a small fraction of E-SFDGs are found in the other filaments ($\sim$16$\%$ for Leo II A, $\sim$11$\%$ for Leo II B, and $\sim$0$\%$ for Leo Minor and Canes Venatici). 

In terms of the characteristics of their morphologies, the E-SFDGs are considered as transitional dwarf galaxies -- those on the way to transforming into red, quiescent dwarf early-type galaxies (dEs) from late-type progenitors -- dubbed blue-cored dEs (dE(bc)s; e.g., \citealt{Lisker2006}). It has been known that dE(bc)s are mostly found in the outskirt of clusters and moderately dense groups \citep{De_Rijcke2003,Cellone2005,Lisker2006,Tully2008,De_Rijcke2013,Pak2014,Chung2019}. 

Ram pressure stripping and galaxy harassment have been proposed as the two most discussed formation processes of dE(bc)s in a cluster \citep{Moore1998,Mastropietro2005,Lisker2006}. However, the ram pressure stripping would be nearly absent in a filament environment where no observational evidence for the detection of a hot intrafilamentary medium with high density has not been reported (but see \citet{Benitez-Llambay2013} for a simulation of cosmic web stripping). Galaxy harassment occurs in the central region of a cluster via multiple gravitational interactions with massive member galaxies. However, this process could also be less relevant in the Virgo filaments where the galaxy number density is relatively low, and most of the galaxies consist of faint low-mass dwarf ones.

In the case of the group environment with relatively low-velocity dispersion, gravitational tidal interactions are suggested as the most probable mechanism for the formation of dE(bc)s (see \citealt{Pak2014} for details of the Ursa Major group where direct evidence of galaxy interaction is reported). Regarding the interaction probability, it is noteworthy that the median $f$ value of early-type galaxies in the Ursa Major group ($-$3.22; \citealt{Pak2014}) is comparable to that of the E-SFDGs in the Virgo III filament ($-$3.78) but higher than E-SFDGs in the other filaments (e.g., $-$4.54 for Leo II A and $-$4.49 for Leo II B filaments). Furthermore, the Virgo filaments are known to contain a large number of groups in their structures \citep{Lee2021,Castignani2021}. Consequently, we suggest that the local environment of the Virgo III filament may be more similar to that of the group compared to the other Virgo filaments, where a high frequency of transitional dwarf galaxies can emerge through frequent galaxy interactions.

Another possible mechanism for the formation of dE(bc)s is the merger between low-mass galaxies with available gas reservoirs (e.g., \citealt{Bekki2008,Watts_Bekki2016}). The mergers between galaxies lead to the triggering of starburst in the central region through the infall of gas, which eventually forms a galaxy morphologically similar to dE(bc)s. The observed properties of the internal kinematics of dE(bc)s in the Virgo cluster is evidence that a merging event is a formation channel for at least some of the transitional dwarf galaxies \citep{Chung2019}. In contrast to the cluster environment, galaxy mergers are more frequent in low-density environments such as the galaxy group with low relative velocities between galaxies \citep{Binney_Tremaine1987}. Therefore, some of E-SFDGs in the Virgo filaments with local environment comparable to that of group could be formed by this event. In this respect, studies of the internal kinematics of large samples of E-SFDGs in the Virgo filaments are highly expected.

\section{Summary and Conclusions}
In this paper, we presented the chemical properties and star formation activity of SFDGs in five filamentary structures physically connected to the Virgo cluster by comparing them with those in the Virgo cluster and field. Our main results are as follows:
\\
\\
(1) We selected 210, 241, and 305 SFDGs in the Virgo filaments, Virgo cluster, and field, respectively, using the BPT diagram constructed from the SDSS spectroscopic data. The SFDGs were divided into two subsamples of E-SFDGs and L-SFDGs based on their morphologies classified by the visual inspection of SDSS images. We derived gas-phase metallicities (i.e., O/H) and NUV SFRs of SFDGs using the SDSS spectroscopic and GALEX UV photometric data, respectively.

(2) We found that SFDGs in the Virgo filaments show different distributions of O/H vs. stellar mass and sSFR vs. stellar mass from those in the Virgo cluster. The SFDGs in the Virgo filaments deviate toward lower O/H values and higher sSFRs than those in the Virgo cluster at a given stellar mass. However, SFDGs in the Virgo filaments exhibit similar distributions to those in the field without showing a significant systematic difference. 

(3) The majority of SFDGs in the Virgo cluster is located outskirts (i.e., $>$ R$_{vir}$) of the cluster. Two subsamples of SFDGs in the Virgo cluster located inside and outside of one R$_{vir}$ of the Virgo cluster show no distinct difference in the O/H vs. stellar mass and sSFR vs. stellar mass distributions. These results suggest that ram pressure does not play an essential role in the chemical properties and star formation activity of SFDGs in the Virgo cluster. 

(4) We inspected perturbation parameter $f$ value distributions of SFDGs in the Virgo filaments and Virgo cluster, which provide insights into the effectiveness of interactions between galaxies and local environments of SFDGs. We found that SFDGs in the Virgo cluster exhibit a skewed distribution toward higher $f$ values, whereas SFDGs in the Virgo filaments have a lower peak. This lends to the proposal that tidal interactions between galaxies could be a possible primary process in the metallicity enhancement and star formation suppression of SFDGs in the Virgo cluster. 

(5) The SFDGs in the Virgo III filament appear to have O/H and sSFR distributions similar to those of the Virgo cluster, whereas SFDGs in the other four filaments show field-like distributions. We also found that the $f$ value distribution of SFDGs in the Virgo III filament is remarkably similar to that of the Virgo cluster, where a large fraction of SFDGs have high $f$ values. We propose that SFDGs in the Virgo III filament are in a different local environment from those in the other four filaments, but they are somewhat comparable to the Virgo cluster where the tidal interactions between galaxies occur more frequently. The similarity of chemical properties and star formation activities of SFDGs in the Virgo III filament with their counterparts in the Virgo cluster indicates that a significant fraction of SFDGs in the Virgo III filament may have already been "chemically" pre-processed before they fell into the Virgo cluster. 

(6) We found that E-SFDGs are chemically evolved objects with low active star formation in the Virgo filaments and Virgo cluster. The O/H and sSFR distributions of E-SFDGs in the Virgo III filament are similar to those of the Virgo cluster. Moreover, about half of SFDGs in the Virgo III filament are E-SFDGs, whereas other filaments contain very small fractions of E-SFDGs. We propose that these E-SFDGs are transitional dwarf galaxies on the way to transforming into red, quiescent dEs from late-type galaxies. The tidal interactions between galaxies in the filaments could be the most probable formation mechanism of these galaxies. We also suggest that galaxy mergers may be an additional formation channel for at least some E-SFDGs in the filament.

Consequently, we emphasize that SFDGs in different local environments are well separated with respect to their properties. The SFDGs associated with tidal interactions are more evolved transitional objects characterized by chemical enhancement, suppressed star formation, and early-type morphologies. Therefore, we propose that pre-processing depends mainly on the small-scale, local environments of the filaments which is in line with the suggestion of local galaxy density as the main parameter of pre-processing \citep{Castignani2021}.\\

We are grateful to the anonymous referee for helpful comments and suggestions that improved the clarity and quality of this paper. J.C., acting as the corresponding author, acknowledges support from the Basic Science Research Program through the National Research Foundation (NRF) of Korea (2018R1A6A3A01013232). S.C.R., acting as the corresponding author, acknowledges support from the Basic Science Research Program through the NRF of Korea (2018R1A2B2006445). S.K. acknowledges support from the Basic Science Research Program through the NRF of Korea (2019R1I1A1A01061237). Y.D.L. acknowledges support from Basic Science Research Program through the NRF of Korea (2020R1A6A3A01099777).\\
\clearpage
%-----------------Table-------------------

\begin{table*}
\caption{Summary of Sample Galaxies}  \label{table1}
\centering 
\begin{tabular}{l r r r r} 
\hline
\hline 
Name  &    Total number of galaxies    & Number of dwarf galaxies    & Number of SFDGs  & Number (Fraction) of E-SFDGs    \\ 

\hline % inserts single horizontal line

Virgo III Filament      & 181        & 76   & 40    & 18 (45$\%$)  \\
Leo Minor Filament      & 54         & 44   & 20    & 1 (5$\%$) \\
Leo II A Filament       & 180        & 121  & 83    & 13 (16$\%$) \\
Leo II B Filament       & 105        & 85   & 45    & 5 (11$\%$) \\
Canes Venatici Filament & 51         & 35   & 22    & 1(5$\%$) \\
Virgo Cluster           & 1589       & 1325 & 241   & 64 (27$\%$) \\
Field                   & 505        & 336  & 305   & 35 (12$\%$) \\

\hline
\end{tabular}
\end{table*}

%-----------------Figure-------------------

%-----------------Figure---------------------

%-----------------Figure-------------------

%-----------------Figure-------------------

%-----------------Figure-------------------

%-----------------Figure-------------------

%-----------------Figure-------------------

%-----------------Figure-------------------

%-----------------Figure-------------------

%-----------------Figure-------------------

\clearpage
\end{document}